\documentclass[runningheads,orivec]{llncs}
\usepackage[T1]{fontenc}
\usepackage{graphicx}
\usepackage[numbers,sort&compress]{natbib}
\usepackage{amsmath,amssymb,amsfonts}
\usepackage{algorithmic}
\usepackage{graphicx}
\usepackage{textcomp}
\usepackage{comment}
\usepackage{microtype}
\usepackage{xcolor}
\usepackage{tabularx}
\usepackage{array} 
\usepackage{multirow} 
\usepackage{colortbl} 
\usepackage{graphicx} 
\usepackage{pifont} 
\usepackage{subcaption}
\usepackage[misc,geometry]{ifsym}

\usepackage[hyphens]{url}

\usepackage[obeyFinal]{todonotes}

\usetikzlibrary{positioning,fit,arrows.meta,backgrounds}

\begin{document}
\title{AUTOPSY: A Framework for Tackling Privacy Challenges in the Automotive Industry}
\titlerunning{AUTOPSY: A Framework for Privacy in the Automotive Industry}
%
\author{Sebastian Pape \textsuperscript{\Letter} \inst{1,2}\orcidID{0000-0002-0893-7856} \and
Anis Bkakria\inst{3}\orcidID{0000-0002-9758-4617} \and
Maurice Heymann\inst{1}\orcidID{0009-0008-5603-8653} \and
Badreddine Chah\inst{4}\orcidID{0009-0003-2665-8679} \and
Abdeljalil Abbas-Turki\inst{4}\orcidID{0000-0002-4443-0542} \and
Sarah Syed-Winkler\inst{1}\orcidID{0000-0002-3197-9456} \and
Matthias Hiller\inst{5}\orcidID{0000-0003-1238-1114} \and
Reda Yaich\inst{3}\orcidID{0000-0001-7294-5909}}

\institute{Continental Automotive Technologies GmbH, Frankfurt, Germany, \email{firstname.lastname@continental.com}
\and
Goethe University Frankfurt, Germany
 \and
IRT SystemX, Palaiseau, France
\and
CIAD UR 7533, UTBM, Belfort, France
\and
Fraunhofer AISEC, Garching, Germany
}

\authorrunning{S. Pape et al.}
%
\maketitle              
\begin{abstract}
With the General Data Protection Regulation (GDPR) in place, all domains have to ensure compliance with privacy legislation. However, compliance does not necessarily result in a privacy-friendly system as for example getting users' consent to process their data does not improve the privacy-friendliness of the system. Therefore, the goal of the AUTOPSY project was to support the privacy engineering process in the automotive domain by providing several building blocks which technically improve the privacy-friendliness of modern, i.\,e., connected and (partially) automated vehicles.

This paper presents the results of the AUTOPSY project: a system model to identify relevant entities and locations to apply privacy enhancing technologies (PETs); the privacy manager aiming at more control of the data flow from the vehicle, a PET selection approach based on GDPR principles, and an architectural framework for automotive privacy. Furthermore, we built a demonstrator for location-based services to evaluate the architectural framework.

\keywords{privacy engineering\and automotive privacy\and pet selection\and privacy-enhancing technologies\and demonstrator}
\end{abstract}
\section{Introduction}\label{sect:introduction}
Faster and faster innovation cycles in the IT and electronics industry lead to new global trends such as IoT, cloud, or AI that change several aspects of our daily lives. For example, in the transportation domain, trends such as autonomous driving, e-mobility and shared mobility will significantly impact our ways of life towards more safety, quality of life, and sustainability. 
The development and deployment of connected vehicles, smart car applications and automotive services (mobility, traffic and parking management, etc.) relies on the generation, processing and sharing of unprecedented amounts of data. This need is further exacerbated in the context of autonomous driving vehicles wherein data, that needs to be shared with other vehicles, the infrastructure, and remote services hosted in the Cloud, is of particular importance.
On the other hand, the automotive industry has also a track of extensive privacy policies~\cite{PTKSP18ifipsec} as recently investigated by the Mozilla foundation\footnote{\url{https://foundation.mozilla.org/en/privacynotincluded/categories/cars/}} and huge data leakages\footnote{cf.\ https://media.ccc.de/v/38c3-wir-wissen-wo-dein-auto-steht-volksdaten-von-volkswagen}
, especially of location data.

From the regulatory side, the General Data Protection Regulation (GDPR)~\cite{gdpr} is a huge step forward. However, often companies dealing with data are afraid of being incompliant and see the regulations as a hurdle that needs to be overcome instead of a baseline that represents the absolute minimum for privacy~\cite{ABMPPB24iwpe}. However, compliance with the data protection regulation does not automatically
equal ethical organizational procedures~\cite{tronnier2022discussion} or results in a privacy-friendly system: I.\,e., having a customer sign an end user license agreement that grants the right to process sensitive data, does not solve the underlying issue of actually protecting the data.
Therefore, there is the need to ensure privacy on a technical level: privacy enhancing technologies (PETs) need to be integrated into the real-world information systems engineering process~\cite{pallas2024privacy}.

Focus of this paper is to share the lessons we learned during the project, compare them to results of similar work and derive future work. For that purpose, we briefly introduce the results of the project, which mostly are covered in more detail in the referenced work.

The remainder of the paper is as follows: The remainder of Sect.~\ref{sect:introduction} summarizes the project's goal, contribution and specific use cases. Sect.~\ref{sect:results} describes the results of the AUTOPSY project and Sect.~\ref{sect:evaluation} its evaluation. Sect.~\ref{sect:discussion} discusses lessons learned, connects the results to related work and discusses limitations and future work. Sect.~\ref{sect:conclusion} concludes the paper.

\subsection{Goal and Contribution of the AUTOPSY Project}
The goal of the AUTOPSY project was to provide technical solutions improving the privacy engineering processes of systems in the automotive domain. Our focus was on practical applicability in real-world information systems~\cite{pallas2024privacy}.

This paper contributes several technical building blocks to improve the privacy of connected and partially autonomous vehicles:
a system model to identify relevant components and places to apply privacy enhancing technologies (PETs); the privacy manager, a system aiming at controlling the data flow from the vehicle; a PET selection approach based on GDPR principles; and a framework for automotive privacy.
\subsection{Use Cases Focused on by AUTOPSY poject}
Our project covers four use cases:
\subsubsection{Usage-Based Insurance (UBI)}

Identifying risky driving behaviors is key to reducing road accidents. This Usage-Based Insurance (UBI) use case focuses on Pay-As-You-Drive Insurance\footnote{UBI Source: \url{https://www.mobilize-fs.com/fr/actualites/pay-you-drive-et-pay-how-you-drive-nos-premieres-offres-dassurance-connectee}} type, where premiums depend on driving behavior. We explore how Advanced Driver Assistance Systems (ADAS) detect hazardous driving and use vehicle connectivity to dynamically update driver profiles. However, balancing safety incentives with privacy concerns remains a challenge, as is ensuring accurate driver profiling for fair premium calculation.

Machine learning-based classification is well-suited for this task~\cite{10.1007/978-3-031-66428-1_34}, with federated learning (FL) enabling decentralized model updates without directly sharing user data. However, FL is vulnerable to privacy attacks~\cite{NEURIPS2021_08040837,10.1145/3510032,9546481}.
The key challenge is integrating PETs into FL to maintain both accuracy and privacy, ensuring effective driver classification without exposing sensitive data.

\subsubsection{Platooning}
Platooning \cite{platooning} is a technique close to fully automated driving, where a so-called platoon leader drives a route and permits other vehicles to join the platoon and follow by driving autonomously, which is more convenient and also efficient. The use case requires the vehicles to have a basic set of sensors to sense the environment, cruise control and lane-keep assistance and positioning including a high-definition map.
Leaders announce their routes, and followers privately compute their own, verifying leader trustworthiness before joining. While a server can facilitate this, platooning is possible without one. During operation, V2X messages manage smooth merging, departures, and route deviations. The main privacy challenge is to form a platoon without revealing sensitive information about participants' vehicles.


\subsubsection{Silent Testing}
Silent testing is a testing method that utilizes data from automated vehicles in a customer fleet. \citet{wangsilent} summarizes the key concepts of silent testing. We refer to silent testing as the monitoring of a System under Test (SuT) in real-world conditions on public roads, without controlling the vehicle’s actuators.
A human driver or automated driving system performs the driving task while the SuT is running, but neither the driver nor passengers actively participate in the monitoring. These restrictions ensure that no safety risks are posed. A primary use case of the collected data is safety validation of the vehicle, but it can also contribute to continuous software improvement or provide training data for neural networks. Data collections in silent testing vary significantly depending on the specific monitoring and development goals.\\
The main privacy challenge of this use case is to have meaningful data available for the software improvement or training and validation of machine learning algorithms while limiting the use of that data to derive other insights (not needed for the primary purpose) in parallel.
\subsubsection{Location-Based Service}
Location-Based Services (LBS) provide personalized, location-aware services such as route optimization, nearby recommendations, and contextual information to enhance the user experience. LBS rely on real-time GPS and user preferences to offer timely, relevant data. To deliver accurate services, the LBS provider (LBSP) collects and processes data from users, e.\,g.\ current position (GPS data), direction of movement (to anticipate needs), and user queries (preferences for points of interest like restaurants or parks).\\
The reliance of LBS on continuous real-time location data collection brings inherent privacy challenges for users. The sensitive nature of location information can expose users to risks if mismanaged, including potential tracking of movement patterns and revealing personal routines.

\section{AUTOPSY results}\label{sect:results}
Our results contribute to various areas of privacy engineering. In the design phase our system model is supporting the identification of relevant entities and suitable locations to integrate PETs in modern vehicle architectures (cf.\ Sect.~\ref{sect:result_systemmodel}). In the same manner, the privacy manager (PM) provides an abstract concept to control the application flow inside the vehicle (cf.\ Sect.~\ref{sect:result_privacymanager}). When specific requirements, i.\,e.\ the GDPR principles have been identified, we support also the selection of suitable PETs (cf.\ Sect.~\ref{sect:result_petselection}).
Bringing this altogether, we designed a framework for automotive privacy (cf.\ Sect.~\ref{sect:result_framework}) which we then implemented (cf.\ Sect.~\ref{sect:evaluation_framework}) and instantiated in form of a demonstrator (cf.\ Sect.~\ref{sect:evaluation_demonstrator}) for evaluation.

\subsection{Architectural Concept}
The architectural concept consists of the system model used for analysis, the privacy manager as a concept to enforce data flows, and a proposal how to technically enforce purpose limitation.
\subsubsection{System Model}\label{sect:result_systemmodel}
\begin{figure}[tb]
  \centering
  \includegraphics[width=\linewidth]{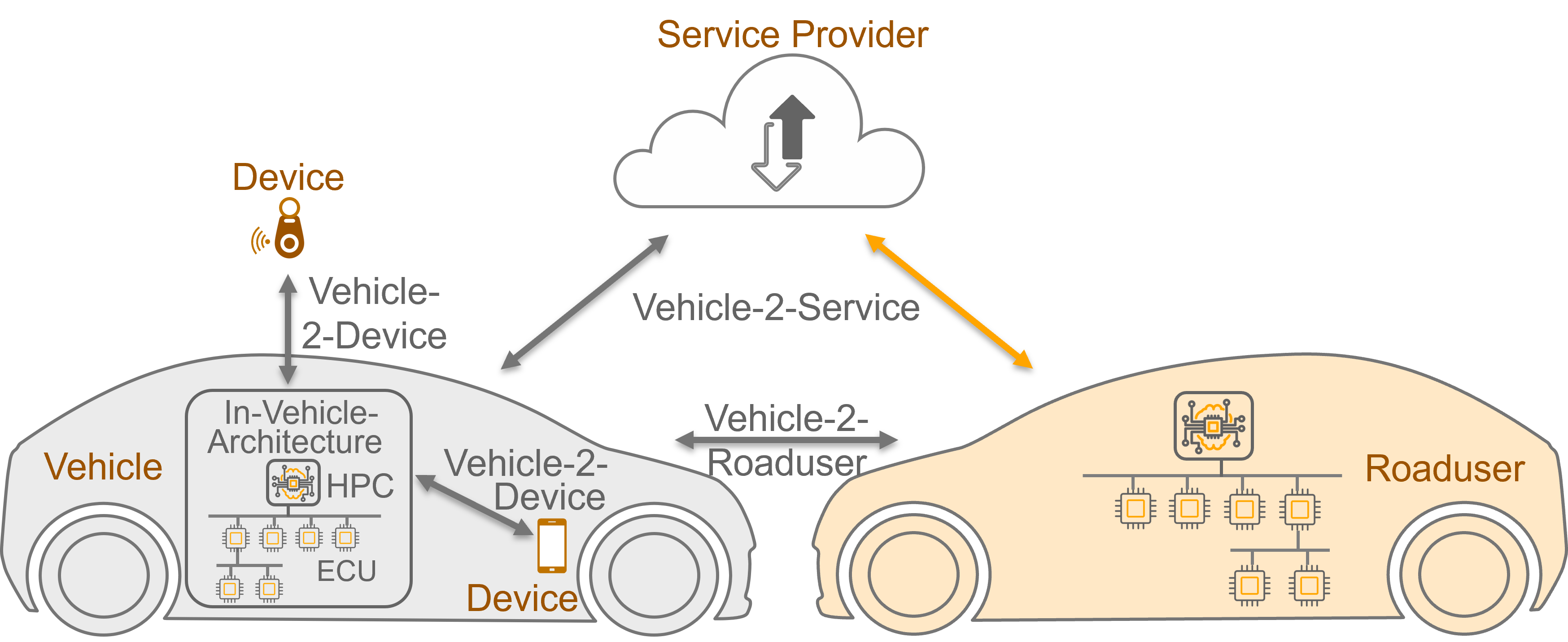}
  \caption{System Model Overview~\cite{PSGCBHWLAY23cscs}}\label{fig:scenario_all}
\end{figure}

We contribute to the field of privacy-by-design in vehicles with a generic system model for the purpose of identifying the relevant entities, data flow and spotting suitable locations to apply PETs. The system model consists of several dimensions to allow capturing different notions: An overview for the relation of the entities with a fine-grained level definition (cf.\ Fig.~\ref{fig:scenario_all}), functional and topological In-Vehicle-Architecture, a communication model, a trust model, and a data classification~\cite{PSGCBHWLAY23cscs}. Our evaluation by applying the system model and the PM to two distinct use cases shows that our approach is feasible and supports the integration of privacy.

\subsubsection{Privacy Manager}\label{sect:result_privacymanager}
A subsequent contribution is the concept of our PM, which supports applications in the implementation of PETs or enforces a certain data flow avoiding that information is leaked
~\cite{PSGCBHWLAY23cscs}.

Fig.~\ref{fig:framework} depicts an abstracted view of the vehicle's architecture. We assume that the PM and potential applications are running in an high-performance computing platform (HPC). There are four different modes in the vehicle for operating with the PM: i) The application has \emph{direct access} to the required data. ii) The application communicates with the PM over an API to \emph{apply PETs} on the required data. iii) The application requests data from the PM, which only delivers the result of a \emph{computation on raw data} and not the raw data itself. iv) A \emph{combination} of PET application and computation on raw data.


\subsubsection{Technically Enforcing Purpose Limitation}\label{sect:results_purpose}
There are two components to the principle of purpose limitation: i) personal data must only be collected for specified, explicit, and legitimate purposes; ii) the collected data cannot be used for a different or incompatible purpose. 

The use of encryption is one technical way to restrict access only to those entities who process data for the specified purpose: Only entities with legitimate purposes are able to decrypt (selected) personal data to prevent entities who do not process data for the stated purpose from processing. This allows the controller to monitor if entities receive the decryption key, which enables them to access specific data. We propose a method how to technically enforce purpose limitation using Attribute-Based Encryption (ABE)~\cite{SPS22cscs}.

\subsection{Supporting PET Selection}\label{sect:result_petselection}
\begin{figure}[tb]
  \centering
  \includegraphics[width=.75\linewidth]{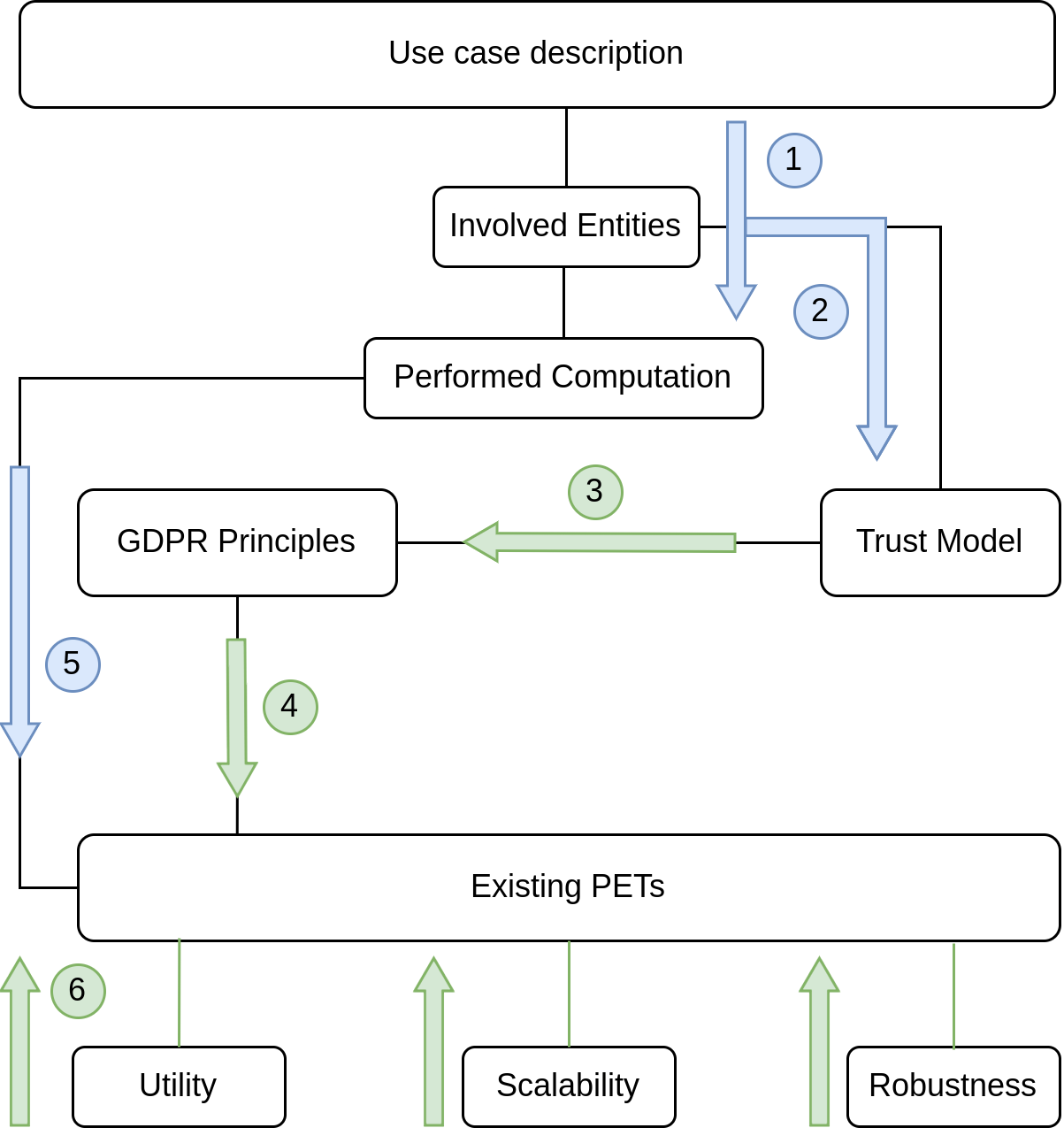}
  \caption{PET Selection Overview~\cite{PBCHS25ares}}\label{fig:PET_Selection_Overview}
\end{figure}

We support the ability to select candidate PETs tailored to specific use cases based on GDPR principles~\cite{gdpr}. Fig.~\ref{fig:PET_Selection_Overview} provides an overview of the necessary steps. As preparation (blue steps, outside of the scope of our framework),
the use case needs to be described, involved entities need to be identified along
with the established trust model. The next steps (in green) are in the focus of our
framework: Based on the trust model, relevant GDPR principles are identified
(step 3, cf. Sect.~\ref{sect:result_petselection_1}). The GDPR principles are then mapped to PET types (step
4, cf. Sect.~\ref{sect:result_petselection_2}) to identify candidate PETs that can be applied to enforce the selected
principles. Furthermore, we provide an assessment of the maturity of PETs with
regards to utility, scalability, and robustness to support the PET selection process
(step 6, cf. Sect.~\ref{sect:result_petselection_3}). Finally, trade-offs between utility, privacy, and efficiency need to be done specifically for the use case (outside of scope of our framework).

\subsubsection{Trust Model vs. GDPR principles}\label{sect:result_petselection_1}
We considered two dimensions: \textit{i) Relevance} of the principle, which does not imply that there are not relevant principles, but that in each trust model some may be more relevant than others. \textit{ii) Accountability} which refers to the entity's capability to not only follow the principle but also provide some evidence.

\subsubsection{Mapping GDPR principles to PETs}\label{sect:result_petselection_2}


PETs apply across four layers: \textbf{Physical} (vehicles with sensors and communication), \textbf{Communication} (data transmission via DSRC and cellular networks), \textbf{Processing} (processing of data inside and outside the vehicle), and \textbf{Storage} (storage of data, in particular, outside of the vehicle). Choosing PETs requires mapping GDPR principles to these layers (see Table \ref{GDPR_To_PETs}). In Table \ref{GDPR_To_PETs}, the green check indicates that most PET types effectively address this specific GDPR principle, while the yellow check suggests that their applicability may vary depending on the utilization context and use case. Compliance cannot rely on a single layer; each contributes by implementing and combining PETs accordingly.

    
    


The physical layer uses PETs like Attribute-Based Encryption (ABE), Local Differential Privacy (LDP), and pseudonymisation to support GDPR principles such as purpose limitation, data minimisation, and confidentiality. 
The communication layer relies on authentication (i.e., digital signatures), encryption, and anonymity for data minimisation, integrity and confidentiality, and accountability. The processing layer is the most interesting layer for using PETs to address GDPR principles like purpose limitation and data minimisation. Furthermore, Trusted Execution Environments (TEE) can be used to protect the confidentiality and integrity of processed data. In the storage layer, it can also be used to ensure deletion and address storage limitation, though it depends on the service provider's hardware control. Additionally, encryption can be used, i.e., ABE allows the enforcement of purpose limitation.

\begin{table}[tb]
\centering
\begin{tabularx}{\columnwidth}{|l|l|X|X|X|X|X|X|X|}
\hline
\textbf{Layer} & \textbf{PET Type} & \rotatebox{90}{\textbf{L/F/T}} & \rotatebox{90}{\textbf{PL}} & \rotatebox{90}{\textbf{DM}} & \rotatebox{90}{\textbf{A}} & \rotatebox{90}{\textbf{SL}} & \rotatebox{90}{\textbf{I/C}} & \rotatebox{90}{\textbf{Acc}} \\ 
\hline
\hline
\multirow{2}{*}{\textbf{Physical}} 
    & Anonymity-based &  &  & \cellcolor{green!25} \checkmark &  &  &  &   \\\cline{2-9}
    & Cryptography-based &  & \cellcolor{green!25} \checkmark & \cellcolor{green!25} \checkmark & &  & \cellcolor{green!25} \checkmark &     \\ 
\hline
\hline
\multirow{3}{*}{\textbf{Communication}} 
    & Anonymity-based &  &  & \cellcolor{green!25} \checkmark &  &  &  &    \\ \cline{2-9}
    & Authentication-based &  &  &  &   &  &  \cellcolor{green!25} \checkmark  & \cellcolor{green!25} \checkmark      \\ \cline{2-9}
        & Cryptography-based &  &  &  &  & & \cellcolor{green!25} \checkmark  &    \\ \cline{2-9}
\hline
\hline
\multirow{3}{*}{\textbf{Processing}}  
    & Anonymity-based &  &  & \cellcolor{green!25} \checkmark &  &  &  &   \\ \cline{2-9}
     & Cryptography-based & \cellcolor{yellow!25} \checkmark & \cellcolor{green!25} \checkmark & \cellcolor{green!25} \checkmark & &  & \cellcolor{green!25} \checkmark &      \\ \cline{2-9}
    & Traceability & \cellcolor{yellow!25} \checkmark & \cellcolor{yellow!25} \checkmark & \cellcolor{green!25} \checkmark &  &  &  & \cellcolor{yellow!25} \checkmark  \\ 
\hline
\hline
\textbf{Storage}  
    & Cryptography-based &  & \cellcolor{green!25} \checkmark  &  \cellcolor{green!25} \checkmark &  & \cellcolor{yellow!25} \checkmark & \cellcolor{green!25} \checkmark &    \\\cline{2-9} 
    & Immutability & \cellcolor{green!25} \checkmark &  &  &  &  & \cellcolor{green!25} \checkmark &  \cellcolor{green!25} \checkmark \\
\hline
\multicolumn{9}{c}{}\\
\end{tabularx}
\caption{Mapping of GDPR (Article 5) Principles to PET Types.\\\textbf{L/F/T}: Lawfulness, Fairness, and Transparency; \textbf{PL}: Purpose Limitation;\\\textbf{DM}: Data Minimization; \textbf{A}: Accuracy; \textbf{SL}: Storage Limitation\\\textbf{I/C}: Integrity and Confidentiality; \textbf{Acc}: Accountability }
\label{GDPR_To_PETs}
\end{table}
\subsubsection{Maturity comparison}\label{sect:result_petselection_3}
We performed a comparative analysis of existing PETs, revealing their effectiveness, robustness, scalability, and suitability for low-computation power devices.
The assessed maturity level can be used to decide about the practical applicability for the shortlisted PETs. Our analysis identified key PETs, such as LDP \cite{yang2024local} and FL \cite{mcmahan2017communication} as optimal choices for our use cases due to their adaptability. 

\subsection{Automotive Privacy Framework}\label{sect:result_framework}

Across the studied use cases, a \textbf{Three-party Model} emerges as an adaptable framework for privacy-preserving data processing:

\begin{itemize}
    \item \textbf{Vehicle / User} – The end-user who selects and installs applications or services requiring access to vehicle data.
    \item \textbf{Third party: intermediate servers} – Collect and perform data processing without direct access to private information, leveraging, e.\,g., encryption and secure multi-party encryption.
    \item \textbf{Service Providers} – Utilize data provided by intermediate servers to deliver services.
\end{itemize}

This structure is similar to the Distributed Aggregation Protocol (DAP) \cite{ietf-ppm-dap-14}, which can be considered a three-party model where aggregators (i.\,e., intermediate servers) process data securely. Like DAP, strong security requirements are imposed on intermediate servers, ensuring they do not collude with each other, the service provider, or malicious users.

However, notable differences in data processing emerge between distinct use cases, such as data aggregation, personalized computations, credential issuance, verification, etc. These differences highlight the challenge of the lack of a unified protocol that fully addresses the specificities of different use cases, despite the structural similarities among the models.

Within the Automotive system, there are several key entities that interact to ensure secure and privacy-preserving data access for applications (cf.\ Fig.~\ref{fig:framework}):

\begin{figure}[tp]
  \centering
  \includegraphics[width=\textwidth, clip, trim={.75cm .75cm .6cm .75cm}]{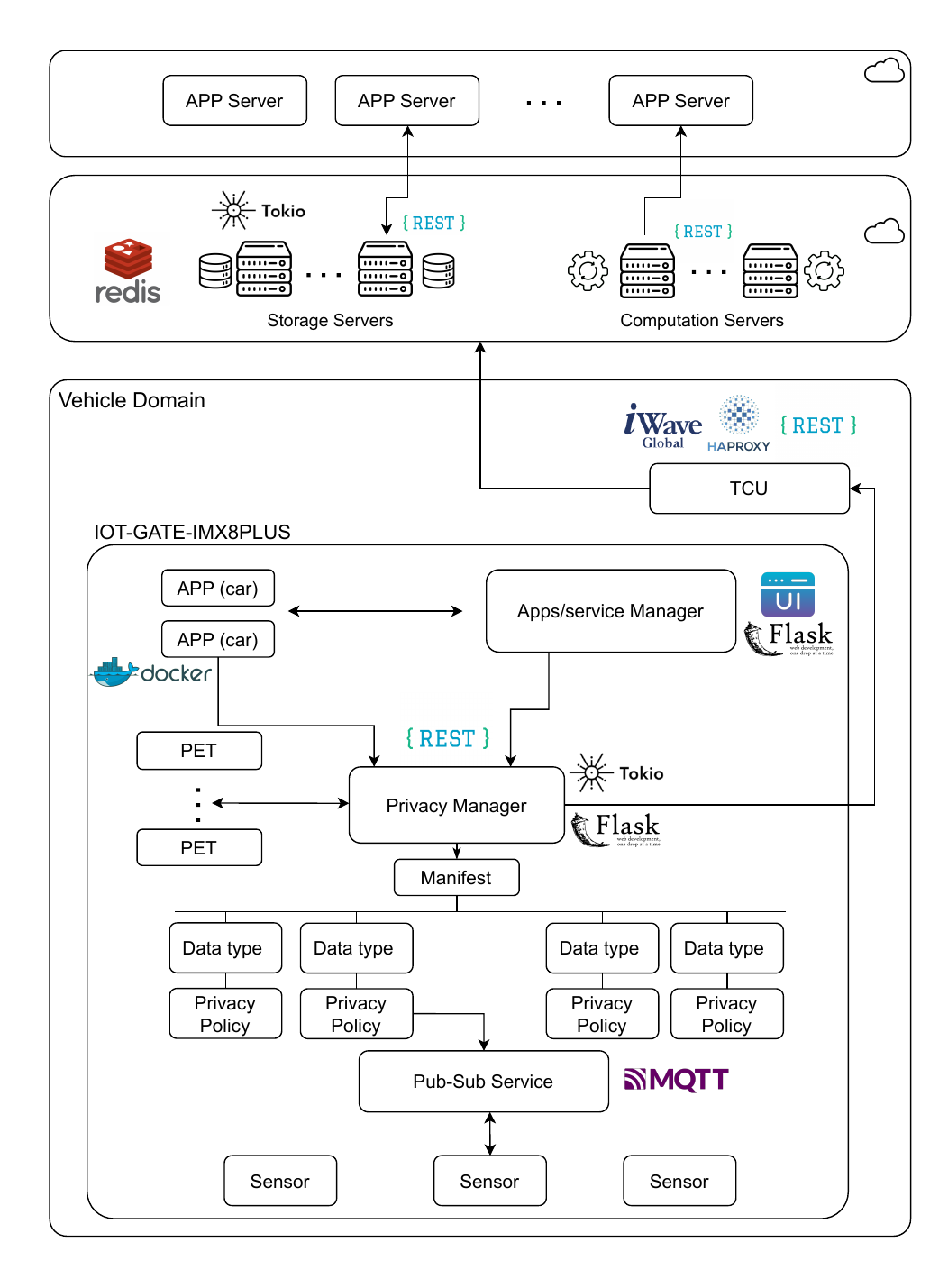}
  \caption{Framework Architecture}\label{fig:framework}
\end{figure}

\begin{itemize}
    \item \textbf{App Manager}: Responsible for retrieving the selected application and extracting its \textit{manifest}, which contains metadata about the data types required by the app.
    \item \textbf{Manifest}: Contains information about the data an App requires, which PETs are supported, and the data constraints such as the precision (e.g., resolution of location, amount of noise added to temperature etc.).
    \item \textbf{Privacy Manager}: A central component that enforces privacy policies. It receives the app’s manifest, determines necessary \textit{PETs} to protect sensitive data, and generates privacy rules that regulate the data flow per data type.
    \item \textbf{Pub-Sub Service}: Service that allows access to data items from their respective sources, i.e., \textit{sensors, Electronic Control Units (ECUs), and the Telematics Control Unit (TCU)}. These services act as intermediaries to retrieve raw data.
    \item \textbf{Privacy-Enhancing Technologies (PETs)}: Mechanisms applied to data (e.g., anonymization, encryption, differential privacy) to ensure compliance with privacy policies before sharing data with applications. These \textit{PETs} are designed to be generic and OS-independent.
\end{itemize}


To ensure privacy and controlled data access within the vehicle domain, applications follow a structured process. When a user installs an app, the App Manager retrieves and analyzes its \textit{manifest}. The PM analyzes the required data types, supported PETs, and determines privacy measures. In general, ABE is applied to all data for storage and access control, ensuring purpose limitation where only authorized service providers can access it. Later, when an app requests data, the PM applies the relevant PETs to the cached data and transmits it securely, ensuring real-time availability and privacy protection.

While an App can individually send data to the service provider / App as described previously, we also support a data-lake approach by streaming updates to the Storage Servers. The Storage Servers always (as long as network access is given) provide (encrypted) sensor data meeting the service's individual constraints such as freshness, precision etc. These parameters are also defined within the App Manifest that is provided when the App Provider uploads it to an AppStore maintained by the trusted third-party (like an OEM). These data updates are bundled and the transmission has been optimized to reduce the required bandwidth while fulfilling the Apps' constraints. Data stored on the Storage Servers already has PETs applied according to the Manifest, for example one item could represent $ABE(DP(data))$ or $ABE(Mask(data))$. Consequently, multiple Apps that require identical data and utilize the same PETs can share the same data item, thereby reducing the storage costs on the servers and the communication costs for the vehicle.

In addition to this data lake approach, we support the processing of the encrypted data of the car in a privacy-friendly way, similar to DAP. However, in the context of this project, data processing extends beyond simple aggregation measures by incorporating additional techniques such as HE and Differential Privacy (DP).

\section{Evaluation}\label{sect:evaluation}

For the evaluation purposes, we implemented the automotive privacy framework described in the previous section and built a demonstrator for the specific use-case of location-based services (cf.\ Sect.~\ref{sect:evaluation_demonstrator}) on top of it. The use case we could also utilize to test our method supporting PET selection.

\subsection{Implementation of the Automotive Privacy Framework}\label{sect:evaluation_framework}
As described before, we implemented the three distinct parties "vehicle / user", "third party: intermediate servers", and "service providers". Within the implementation of the framework, it was necessary to tailor PETs to our automotive use cases:

For the location navigation use case, we extended basic location-based LDP to manage multifaceted privacy risks, including location, trajectory, and visited points of interest, while optimizing utility in continuous LBS interactions \cite{bkakria2024framework}. In the platooning use case, we build upon Homomorphic Encryption (HE) to design a privacy-preserving protocol for platoon group formation \cite{chah2023h3pc}, ensuring secure and efficient vehicle coordination while maintaining data confidentiality. For the UBI use case, we combine FL with LDP and HE to preserve the confidentiality of sensitive driver information while enabling effective machine learning model training and updates \cite{UBI_25}. This approach ensures privacy without compromising model accuracy and adaptability.  
To technically enforce purpose limitation, we could utilize the ABE implementation
rabe\footnote{url{https://github.com/Fraunhofer-AISEC/rabe}}, a rust library implementing several Attribute Based Encryption (ABE) schemes.

\subsection{Demonstrator for Location-Based Services}\label{sect:evaluation_demonstrator}
To showcase the added value of our implemented framework, we use an application that instantiates a location-based service. This application enables users to query a location-based service for nearby points of interest (POIs) of a specific type, such as restaurants, fuel stations, or other relevant locations. When a query is sent by the user, the LBS service retrieves the user's location from the query and sends the nearby points of interest.

\begin{figure}[!t]
    \centering
    \begin{subfigure}[b]{0.4\linewidth}
        \centering
        \includegraphics[width=\textwidth]{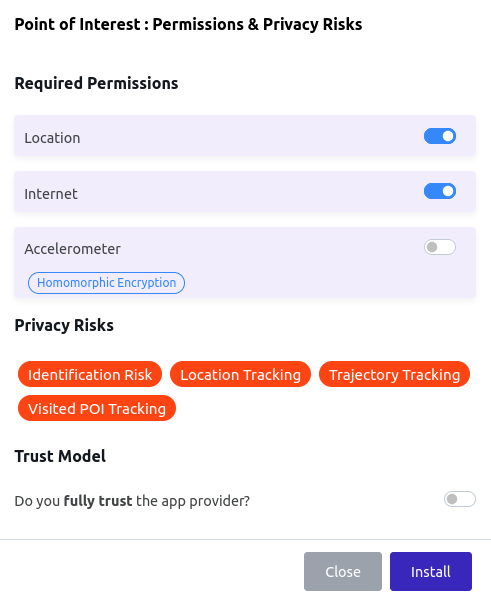}
        \caption{UI of the Location-Based Service App}%
        {{}}    
        \label{fig:app_install}
    \end{subfigure}
    \hfill
    \begin{subfigure}[b]{0.55\linewidth}  
        \centering 
        \includegraphics[width=\textwidth]{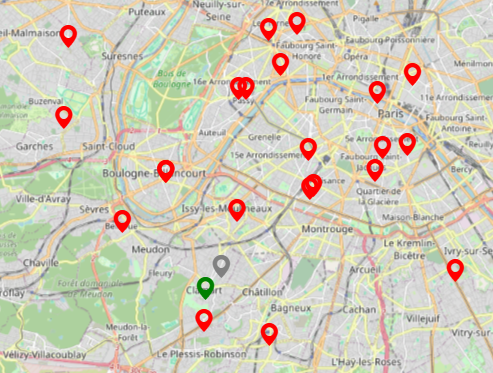}
        \caption{App’s UI Displaying Real and Obfuscated Locations}%
        {{}}    
        \label{fig:locations}
    \end{subfigure}
    \vskip\baselineskip
    \begin{subfigure}[b]{\linewidth}   
        \centering 
        \includegraphics[width=\textwidth]{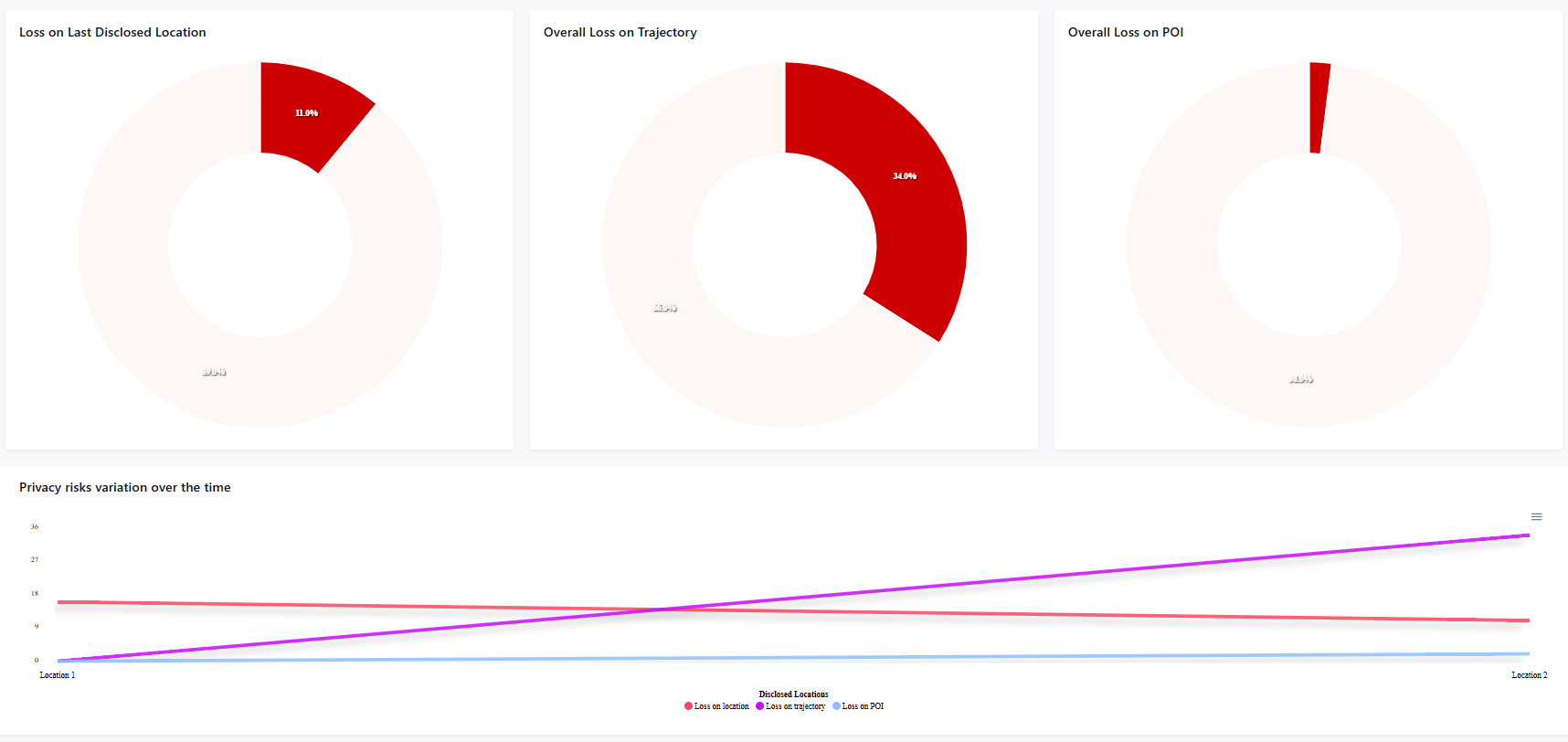}
        \caption{Disclosure's Effect on the Privacy of the User’s Trajectory}%
        {{}}    
        \label{fig:loss_quantification}
    \end{subfigure}
    \hfill
    \caption{Demonstrator's User Interfaces}
\end{figure}

To interact with the LBS, the user must install the LBS app within the car’s system. Based on the requested data types, as depicted in Fig.~\ref{fig:app_install}, the PM evaluates and shows the privacy threats the user may encounter when interacting with the app. Building on these functional requirements and the necessary data set, the PM selects the optimal combination of PETs to achieve the best privacy-utility tradeoff. For the considered LBS application, the PM applies Differential Privacy (DP) using the Planar Isotropic Mechanism \cite{bkakria2024framework} to obfuscate the user's location before sharing it with the LBS app.  

To demonstrate the impact of applying Differential Privacy (DP) to the user's location, the app's UI displays both the real and obfuscated locations, as illustrated in Fig.~\ref{fig:locations} (disclosed location in green, real location on gray, and points of interest are in red). Additionally, for each interaction with the LBS server—where a new obfuscated location is disclosed—the UI visualizes the effect of this disclosure on the privacy of the user's trajectory, as shown in Fig.~\ref{fig:loss_quantification}.

This use case demonstrator shows that our framework effectively balances privacy and utility by leveraging Differential Privacy to protect users' location data while maintaining the functionality of location-based services. By visualizing both real and obfuscated locations, as well as the evolving privacy impact on the user's trajectory, our approach enhances user awareness and control over their data.


\section{Discussion}\label{sect:discussion}
In this section, we first introduce lessons learned through our project, then compare our project results to related work from the privacy engineering domain, and eventually discuss limitations and future work.

\subsection{Lessons learned}

Initially, the integration within the \emph{automotive hardware} was very difficult as we wanted to target a board that was running a partly-proprietary Android-based framework. After switching to another board that supports an ARM build of Debian Linux, the development became significantly easier than expected. As we are now using an ARM Cortex-A53 quad-core CPU with 4GB RAM, we also have a rather high-end build compared to ordinary automotive hardware. Thus, we are rather at the (automotive) high performance computing (HPC) area and the application of PETs has proven to be significantly faster than anticipated, without appearing to create any bottlenecks. However, we had developed the architecture to perform computations whenever possible in a privacy-friendly environment outside of the vehicle.
We noticed that there is a \emph{lack of standardization for PETs}. Especially, when it comes to adding noise to mask raw values, there was no easy way to refer to specific noise levels (e.\,g.\ low/medium/high).

The automotive domain might stand out from other domains as \emph{location data} seems to be by far the most important data type, and in contrast to mobile phones, the user/driver can often not easily disable them.
With connected autonomous vehicles being promised within the next years, there might be a focus shift from \emph{behavioural driving data} towards \emph{support data for autonomous driving} as driving characteristics can no longer be easily fingerprinted in autonomous vehicles.

The diverse set of the studied scenarios has clearly demonstrated the challenges in selecting appropriate cryptographic PETs for specific contexts. For instance, in the real-time platooning scenario, the computational overhead associated with acceleration command encryption posed a significant challenge to maintaining platoon stability. As a result, it became necessary to couple Order-Preserving Encryption with HE to facilitate the computation of vehicle ordering within the platoon~\cite{chah2023h3pc}. However, exposing the order of vehicles introduces a trade-off, as it partially reveals position information to external parties. This trade-off highlights the complexity of balancing privacy preservation and system performance in real-time applications.
In other words, the \emph{practical deployment of PETs is not always seamless} and often requires substantial adaptation efforts to align with real-world constraints. This issue also raises broader concerns regarding methodological gaps and the lack of standardization in the field of privacy-preserving applications~\cite{BATRA2025100164}. One approach to address these challenges is to propose a privacy-preserving architecture for automotive applications. Such an architecture would serve as a foundation for discussion, enabling continuous improvement based on state-of-the-art research, real-world use cases, and feedback from stakeholders involved in the service ecosystem. Another advantage of such an architecture is the standardization of protocols and PET implementations in the automotive industry.
While certain challenges may be mitigated using existing standards, their effectiveness depends on two key factors:
\begin{itemize}
    \item A comprehensive integration within privacy-preserving applications to ensure their relevance in real-world scenarios.
    \item The extension of existing standards with new functionalities that address emerging privacy challenges and technological advancements.
\end{itemize}

This structured approach aims to bridge the gap between theoretical privacy mechanisms and practical implementation constraints, fostering the development of scalable, efficient, and privacy-compliant services.

\subsection{Comparison of Project Results to Related Work}

In this subsection, we discuss how our findings relate to general findings of privacy engineering papers. We contribute to results with more \emph{practical applicability in real-world information systems}~\cite{pallas2024privacy} by focusing on the automotive domain and in particular, investigating the four specific use-cases and developing tailored PETs, the automotive privacy framework along with a demonstrator for location-based services \emph{in a realistic setting}~\cite{pallas2024privacy}. \citet{gurses2011engineering} and \citet{spiekermann2012challenges} state that engineers find it hard to \emph{translate from abstract, ambiguous privacy requirements to specific technologies and solutions}. While this is in general a hard problem, which has already been partially addressed~\cite{lobner2021comparison,rannenberg2021study}, we could contribute to a subset of the challenge by mapping privacy principles of the GDPR~\cite{gdpr} to PETs. The mapping supports privacy engineers by excluding non suitable PETs. The provided assessment of the PETs maturity towards different dimensions then further supports the selection process. However, even though we ease the \emph{PET selection process}, still some specialized expertise will be required~\cite{notario2015pripare}. 

Many approaches focus on the principle of data minimization~\cite{gurses2015engineering}. With our proposal to technically enforce purpose limitation, we are \emph{going beyond anonymization, data minimization and
security}~\cite{pallas2024privacy} and demonstrate that it is possible to address purpose limitation not only by consent and legal work. \citet{kostova2020privacy} highlights the \emph{challenges of dealing with services}. Our automotive privacy framework specifically considers different services from different providers: By bundling outgoing data, we assure that no more data than needed is send to the data lake in the cloud where the data is distributed among the different services. One reason for the architecture decision was also to \emph{reduce performance overheads}~\cite{pallas2024privacy} for applying PETs and the used traffic from vehicles to the cloud backend.

Our framework also supports the \emph{composition of different PETs}~\cite{kostova2020privacy}. In particular, we are applying ABE on top of all other PETs to enforce purpose limitation, and thus also \emph{minimize trust assumptions}~\cite{kostova2020privacy}.

Last but not least, we provide \emph{technical artifacts that are easily reusable in real-world environments}~\cite{pallas2024privacy}: a source code\footnote{\url{https://github.com/badr007-01/Building-a-Database-of-Simulated-Driver-Behaviors-Using-the-SUMO-Simulator}} that runs the SUMO simulator and generates a dataset of three different driver behaviors (Slow, Normal, Dangerous). We use this dataset to initially construct the ML model for the UBI use case. Additionally, we provide a library\footnote{\url{https://github.com/badr007-01/Usage_Based_insurance_DP_HE_from_SUMODataset}} that offers a decentralized ML algorithm combining DP and HE to protect sensitive data .

\subsection{Limitations}
Our technical enforcement of purpose limitation suffers from the limitation that when using ABE, the access to certain data can only be limited to a certain party. If the party is not at least semi-honest, it could use the data for other purposes. To prevent it, sophisticated watermarking or digital right management systems would be needed. However, since our proposal is an improvement of the status quo, we refer to \citet{pallas2024privacy} and relax on so far predominant “all-or-nothing” aspirations.

Additionally, we observed that revocation challenges have not been tackled in our solution yet. Once the personal data is available to a service provider or stored in the data lake, it is challenging~-- if not impossible~-- to revoke already distributed data.

Furthermore, the introduction of a trusted third-party that issues all the keys for \textit{ABE}, makes it effectively becoming the GDPR data controller, meaning that the liability for ensuring correct access control might be also passed to that party. This could introduce regulatory burdens and legal risks.

As a central element of our work is the application of PETs to ensure data privacy and the design of a future-proof Automotive data architecture, it is necessary to point out that many \textit{PETs} are not safe against sophisticated quantum-computers as many \textit{PETs} are based on traditional asymmetric schemes such as RSA or ECC. However, \textit{PETs} such as Zero-Knowledge-Proofs based on hash-based signatures or lattice-based FHE remain secure.

Regarding the PET selection, we did not introduce a measure to evaluate the impact of applying the different possible PETs or their combinations~\cite{ABMPPB24iwpe}. Thus, it is also not clear how close the finally selected solution would be to the optimal solution -- only using the minimal amount of data needed to provide the specific service.

\subsection{Future work}
We support the statements by \citet{kostova2020privacy} that standardization is one of the major challenges. This holds not only for PETs themselves, but also for instantiating / configuring / parameterizing them. Although we proposed a way to coordinate the application of PETs within the vehicle by the privacy manager (PM), the information exchange between the PM and applications needs at least a de facto standard, perhaps similar to the way apps in Android are configured. 

Similar privacy frameworks like ours are needed for other domains as well. This would allow us to answer the question if there are possibilities to generalize the results beyond single domains, such as in our case the automotive domain.

Our work was mostly focused on the privacy architects' and developers' perspective. It would also be interesting to consider the user's perspective, that is, the user acceptance effects and criteria for PETs as already done for machine learning~\cite{LPB23trustbus,LPBP24arxiv}. User perceptions of PETs considering technology use behavior~\cite{HPR20pets} willingness to pay~\cite{HCP19ifipsec}, privacy concerns~\cite{HP18ifipsec}, risk beliefs~\cite{HP19hicss}, privacy literacy~\cite{HP20sigmis}, and perceived anonymity and trust~\cite{HP18amcis} have been investigated in the past by Harborth et al.\ and shown to have a huge influence on the acceptance of PETs by endusers. Therefore, it would be a another challenge to include those into the architecture design and PET selection process, perhaps along with other economic criteria such as licenses of the PETS, fees and operating costs, and availability of knowledgeable experts.

\section{Conclusion}\label{sect:conclusion}
With the AUTOPSY project, we could contribute to the privacy world for automotive with a couple of ideas, such as a system model, the privacy manager, a PET selection process based on GDPR principles, the development of tailored PETs, and the privacy framework resulting in a demonstrator for the location-based service use case.
Given the particular challenges of supply chains in the automotive domain, it will be particular interesting to see if and which of the ideas can make it to real products. Most likely, it will also depend on the efforts and willingness of all players to contribute to standardization of the used PETs, their configuration and the needed interfaces.

\section*{Acknowledgements}
The authors would like to thank Armando Miguel Garcia and Mario Hoffmann for their contributions to the project. This work was suported by the Federal Ministry of Education and Research, Germany (BMBF) under grant number 16KIS1382 and by the Agence Nationale de la Recherche, France (ANR) under grant number ANR-20-CYAL-0008.

\bibliographystyle{splncs04nat}
\bibliography{references}
All URLs have been last accessed on May 14th, 2025.
\end{document}